\documentstyle[epsfig,subfig,12pt]{article}

\protect
\protect
\newcommand{\lgm}{{\,\rm ln }}

\newcommand{\ice}[1]{\relax}
\newcommand{\prd}{\partial}
\newcommand{\ep}{\epsilon}
\newcommand{\beq}{\begin{equation}}
\newcommand{\eeq}{\end{equation}}
\newcommand{\bea}{\begin{eqnarray}}
\newcommand{\eea}{\end{eqnarray}}

\newcommand{\re}[1]{(\ref{#1})}

\def\bbuildrel#1_#2^#3%
{\mathrel{\mathop{\kern 0pt#1}\limits_{#2}^{#3}}}

\newcommand{\ba}{\begin{array}} 
\newcommand{\ea}{\end{array}} 
 
\newcommand{\as}{\alpha_s}

\newcommand{\G}{\Gamma}
\newcommand{\g}{\gamma}

\newcommand{\dmu}{\mu^2\frac{d}{d\mu^2}}
\newcommand{\msbar}{\overline{\mbox{MS}}}
\newcommand{\dsp}{\displaystyle}

\begin{document}  
%**end of header

\begin{titlepage}
%%%
\begin{flushright}
\begin{tabular}{l}
  MPI/PhT/96-117\\
  hep-ph/9610531\\
  October   1996   
\end{tabular}
\end{flushright}
%August   1996   
%\hfill  {\small MPI/PhT/96 - 61}
\mbox{}
\protect\vspace*{.3cm}

%\vspace{-1.0cm}
\begin{center}
  \begin{Large}
  \begin{bf}
Four and Three Loop Calculations in QCD: Theory and Applications${}^*$
 \\ \end{bf} 
\end{Large} \vspace{1cm}
K.G.~Chetyrkin $^{a,b}$
\begin{itemize}
\item[$^a$]
 Institute for Nuclear Research,   
 Russian Academy of Sciences,   \\
 60th October Anniversary Prospect 7a 
 Moscow 117312, Russia 
\item[$^b$]{%
Max-Planck-Institut f\"ur Physik, Werner-Heisenberg-Institut, \\ 
F\"ohringer Ring 6, 80805 Munich, Germany}
\end{itemize}
\vspace{0.2cm}

  \vspace{0.5cm}
  {\bf Abstract}
\end{center}
\begin{quotation}
\noindent
The talk briefly   reviews   the current state of the art in  doing 
multiloop QCD calculations in a completely analytic way. 
In particular, we discuss  recent  analytical calculations of
4-loop  ${\cal O}(\as^3)$ gluino contribution  to $R(s)$
and its implications  to $\Gamma_{tot}(Z \to \mbox{hadrons})$
and $\Gamma (\tau^- \to \nu_\tau + \mbox{hadrons})$;
\end{quotation}
%\vfill

\noindent

\medskip

\medskip

\medskip
\vfill

\noindent 
${}^*$Invited talk  at the 
Cracow International Symposium on Radiative Corrections
1-5 August 1996, Cracow, Poland.
\end{titlepage}

\section{\large\bfseries  Introduction}
The firm knowledge of radiative corrections (r.c.) is a must in
confronting the Standard Model with experimental data. While the
electroweak r.c. are usually required to be known in the leading and
next-to-leading approximations, this is certainly not always the case
for the strong interaction ones. The current (relatively large) value
of the strong coupling constant
$\alpha_s(M_Z)= 0.1202 \pm 0.0033$
\cite{Blondel} 
and the remarkably high precision achieved in high energy $e^+ e^-$
experiments make the latter sensitive to really higher order QCD r.c.
such as ${\cal O}(\as^3)$ contributions to $R(s)$ and $\Gamma_{tot}(Z
\to \mbox{hadrons})$.
% (see , e.g. a recent review \cite{review}).

The talk briefly reviews the current state of the art in doing
multiloop QCD calculations in a completely analytic way.  It is {\em
not} intended as a sort of exhaustive review of the vast subject of
the QCD r.c.  We deal  with analytical calculations of 3- and
4-loop r.c.  to 2-point correlators of bilinear quark currents. Even
more, only  massless correlators are considered (high energy limit).
The choice is dictated in part by my personal taste and in part by the
the fact that most advanced results (at least as for the number of
loops involved) have been obtained in this particular field. The reader
who is looking  for a broader exposition including leading and next-to-leading
r.c.  should consult numerous reviews
\cite{review}.

\section{ \large\bfseries Feynman Integrals up to  3 loops}
Here  we will discuss  briefly  the  tools
now available to analytically compute massless propagators
in higher orders. We limit ourselves
to these  rather restricted  class
of Feynman integrals
due to the following   reasons:
\begin{enumerate}
\item  Practice shows that in many cases
       the methods of asymptotic expansions of Feynman
       integrals (for  a recent review see \cite{smirnov95}) 
       do  produce results numerically very well
       approximating the exact results when the latter are
       available. These methods  reduce
       initial multi-scale  Feynman amplitudes to
       combinations of  massless propagators and massive tadpoles.

\item  A number of  problems can be eventually 
       reduced to evaluation of massless propagators.  A 
       important example is  the evaluation  of so-called RG-functions 
       (that is beta-functions
       and anomalous dimensions).
\item  Thanks to the intrinsic simplicity of the integrals
       under discussion, --- they depend on only one
       nontrivial   scale: an external momentum
       ---  their  analytical
       evaluation  proves to be feasible  in quite high
       orders in the coupling constant.
       The same simplicity provides the possibility  of
       constructing  {\em regular} algorithms  for
       evaluating these integrals as well as  dedicated
       computer programs allowing to perform the 
       calculations in  a  convenient
       and   automatic way. 
\end{enumerate}
\noindent
For brevity  massless Feynman integrals
depending on exactly one external momentum will
be denoted by  {\em p-integrals}.  
At the moment there are tools to analytically
compute arbitrary one-- two-- and three--loop p-integrals
(see below). 
Fortunately,
in many important cases one is interested only 
in the 
absorptive part of massless two-point correlators. 
In this case 
available theoretical tools  are enough to guarantee
at least {\em in principle} the  analytical calculability
of absorptive part  of an arbitrary 4-loop p-integral (see below). 
\noindent

\vglue 0.2cm                       

\noindent
{\em One-loop p-integrals}

\begin{figure}
 \begin{center}
  \begin{tabular}{ccc}
  \subfigure[]{\epsfig{file=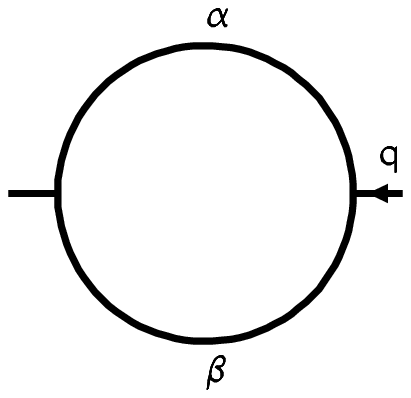,width=3.3cm,height=3.3cm}}
  &
  \subfigure[]{\epsfig{file=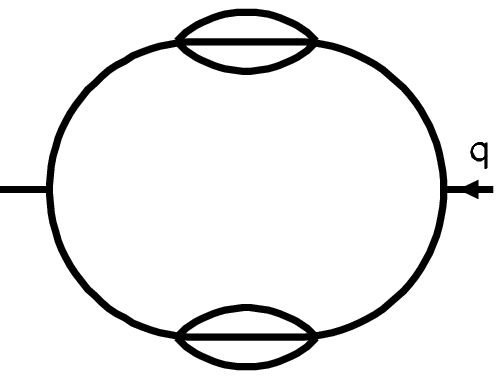,width=3.3cm,height=3.3cm}}
  &
  \subfigure[]{\epsfig{file=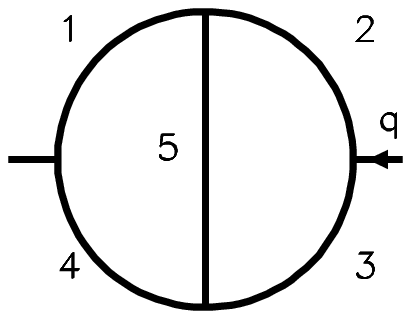,width=3.3cm,height=3.3cm}}
  \end{tabular}
  \end{center}
\caption{
Some p-integrals: (a) the generic one-loop p-integral and
(b) an example of primitive five-loop p-integral;
(c) the master two-loop p-integral. 
 }
\label{Fmethods1}
\end{figure}

We start from a well-known elementary formula for a generic
1-loop p-integral (see Fig.~\ref{Fmethods1}a;
 we shall consider  
Feynman integrals in the Euclidean momentum space throughout
this section)   
\begin{equation}
\begin{array}{c}
\displaystyle
\int 
\frac{{\rm d}^{{\rm D}} \ell }{(2\pi)^{{\rm D}}}
\frac{1}{(q^2)^\alpha (q-l)^{2\beta}} 
= 
\frac{
(q^2)^{2 - \epsilon - \alpha - \beta}}%
{(4 \pi)^{2 - \epsilon}}
G(\alpha,\beta),
\\
\displaystyle
G(\alpha,\beta)  \equiv
\frac{
\Gamma(\alpha + \beta - 2 + \epsilon)}{
\Gamma(\alpha)
\Gamma(\beta)
}
\frac{
\Gamma(2 - \alpha - \epsilon)
\Gamma(2 - \beta - \epsilon)
}{
\Gamma(4 - \alpha - \beta - 2 \epsilon)
}                       
{}\, .                                          
\end{array}
\label{methods1}
\end{equation}
It is of importance to note that any p-integral 
depends homogeneously on its
external momentum. This facts allows the immediate
analytic evaluation   of the whole  class of 
{\em primitive}   p-integrals which, by definition, may be 
performed by repeated application of the one-loop integration 
formula. 
For example, the five-loop scalar
integral of Fig.~\ref{Fmethods1}b  is performed by \re{methods1} 
with  the result
\begin{equation}
\left(
{(q^2)^{-\epsilon}}{(4\pi^2)^{2 - \epsilon})}
\right)^5
(q^2)^{-\epsilon} 
(
G(1,1)
G(1,\epsilon)
)^2
G(1+ 2\epsilon,1+2\epsilon)
{}\, .
\label{methods2}
\end{equation}
\vglue 0.2cm

\noindent
{\em Two-loop p-integrals}

Not all p-integrals are primitive ones. One first
encounters nontrivial p-integrals already
at the two loop level. 
While  one-loop integrals  are  performed with ease
the evaluation of the master  two-loop
diagram (see Fig.~\ref{Fmethods1}c) is not  trivial. 
The corresponding Feynman integral reads 
\begin{equation}
\frac{
( 4 \pi)^{4 - 2 \epsilon}
%(q^2)^{4-2\epsilon-\sum_{{i}} \alpha_{{i}}}
F(\alpha_1,  \dots , \alpha_5 ) 
}{(q^2)^{-4+2\epsilon+\sum_{{i}} \alpha_{{i}}}}
\equiv
\int 
\frac{ \displaystyle
{\rm d}^{{\rm D}} \ell_1
{\rm d}^{{\rm D}} \ell_2} 
{\displaystyle  (2\pi)^{2D} }
\frac{\displaystyle 1}{\displaystyle 
p_1^{2\alpha_1}
p_2^{2\alpha_2}
p_3^{2\alpha_3}
p_4^{2\alpha_4}
p_5^{2\alpha_5}
}
{}\, ,
\label{methods3}
\end{equation}
with the loop momenta
\[
p_1 = \ell_1,       \ \ 
p_2 = \ell_2,       \ \ 
p_3 = q -\ell_2,    \ \ 
p_4 = q -\ell_1,    \ \ 
p_5 = \ell_2 - \ell_1
{}\, . 
\]
A closed expression for
the function $F(\alpha_1,\dots \alpha_5)$ 
for generic values of  the
arguments is not known. However, 
results 
do exist  for
particular cases. The first 
one, valid for a generic space-time 
dimension $D$, was obtained with the 
help of the so-called Gegenbauer 
polynomial  technique in $x$-space (GPTX) \cite{me79}. 
It  reads
\begin{eqnarray}
F(\alpha,1,1,\beta,1)
&{}& 
 =
\frac{G(1,1)}{D -2 -\alpha - \beta}
\label{methods4}
\\ \nonumber
&{}&
\times
\left\{
\alpha[G(\alpha+1,\beta) - G(\alpha+1,\beta +\epsilon)]
+
(\alpha \leftrightarrow \beta)
\right\}
\end{eqnarray}                  
It has been also shown in  Ref.~\cite{me79}
that similar results may be obtained  for
the   case when the indices $\alpha_2, \alpha_3$  and 
$\alpha_5$ are  integers while $\alpha_1$ and $\alpha_4$
are arbitrary.

In practice  one often needs only a few first terms of the
expansion of $F(\alpha_1 \dots \alpha_5)$ in the Laurent series in
$\epsilon$.  This expansion is  known for generic values of the
$\alpha_1 \dots  \alpha_5$  up to a fixed (quite high) order  
(see Refs.~\cite{DKazakov,masterTwoDavid} and  references therein).
\vglue 0.2cm

\noindent
{\em Three-loop p-integrals}

\begin{figure}
\begin{center}
  \begin{tabular}{cc}
   \subfigure[]{\epsfig{file=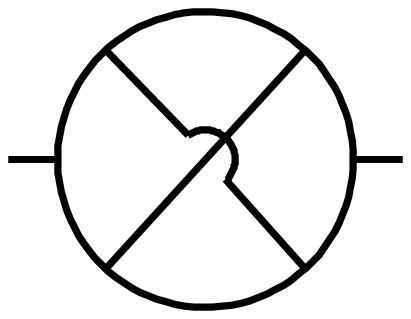,width=4cm,height=4cm}}
   &
   \subfigure[]{\epsfig{file=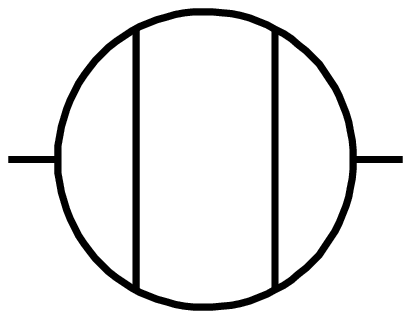,width=4cm,height=4cm}}
   \end{tabular}
\end{center}
\caption{%
(a), (b) the master three-loop
non-planar and planar scalar diagrams.
 }
\label{Fmethods3}
\end{figure}

{\em In principle} GTPX is also applicable to compute some non-trivial
 three-loop p-integrals. For example, the basic scalar non-planar
 three-loop diagram of Fig.~\ref{Fmethods3}a was first calculated via
 GPTX in Ref.~\cite{me79}. However, calculations quickly get clumsy,
 especially for diagrams with numerators.

\begin{figure}
\begin{center}
  \begin{tabular}{ccccc}
  { \Large $\epsilon$ }
  \parbox{2.5cm}{\epsfig{file=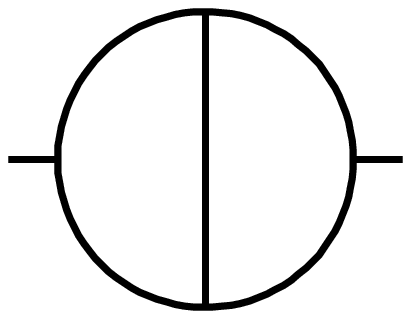,width=2.5cm,height=2.5cm}}
  &$=$&
  \parbox{2.5cm}{\epsfig{file=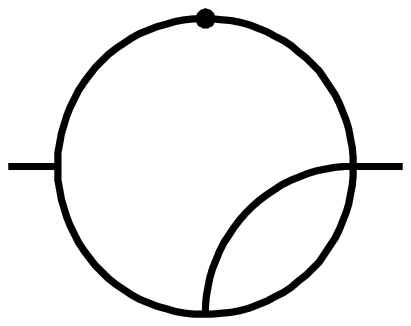,width=2.5cm,height=2.5cm}}
  &$-$&
  \parbox{3cm}{\epsfig{file=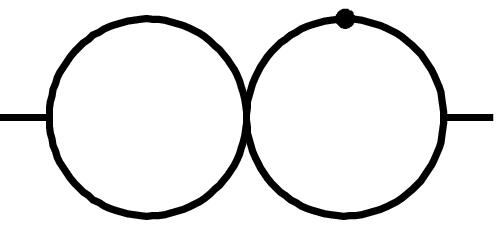,width=3cm,height=3cm}}
  \end{tabular}
\end{center}
\caption{%
The exact relation expressing a nonprimitive two-loop
scalar p-integral through primitive integrals;
a dot on a line means a squared scalar propagator.
 }
\label{Fmethods2}
\end{figure}

The main breakthrough  at the three-loop level happened  with 
elaborating the method of integration by parts  of dimensionally  
regularized integrals Refs.~\cite{me81a,me81b}.  
The key identity for the method  is\footnote{For
two-loop massive integrals a similar identity was used 
in the classical work by ${}'$t~Hooft and Veltman Ref.~\cite{dim.reg-a}.
}
\begin{equation}
\int 
{\rm d}^{{\rm D}} \ell
\frac{\displaystyle \partial}{\displaystyle \partial \ell_\mu}
I(\ell,\dots)
\equiv 0
{}\, ,
\label{methods5}
\end{equation}
where $I(\ell,\dots)$
is a Feynman {\em integrand} and $\ell$ is one of its loop
momenta. 
The identity reflects the  possibility of neglecting the surface terms, 
which holds true in dimensional regularization \cite{me83}.  
The use  of 
\re{methods5} along with  tricks like 
completing momentum squares and 
cancelling  similar factors in the nominator against 
those in the denominator constitutes the essence of the 
approach.  The identity depicted in Fig.~\ref{Fmethods2}
is a typical example of relations obtainable with the 
help of  the integration by parts method. 

It should be, however, heavily stressed that, the validity of such
operations for {\em divergent} dimensionally regulated integrals with
deeply intermixed UV and IR (sub)divergences is {\em not} obvious any
more.  It had to be rigorously justified within a proper
generalization of the dimensional regularization itself.
% {\em per se}. 
It has  been done in Ref.~\protect\cite{me83} (see also a monograph 
\cite{Smi91}).

The general scheme of the use of the 
integration by parts method is based on the
exploitation  the identities of type \re{methods5}
in the form of recurrence relations with the aim
to express  a complicated diagram through the   simpler ones.
All (about a dozen) topologically
different three-loop p-integrals were neatly analyzed 
in Ref.~\cite{me81b}
and a concrete calculational algorithm was suggested for every
topology. As a result the algorithm of integration  by parts 
for three-loop  p-integrals   was developed.
The algorithm  constitutes a series
of involved identities which are used  
to  identically transform any three-loop p-integral
into a sum of primitive one-loop p-integrals and two 
basic three-loop p-integrals pictured in Fig.~\ref{Fmethods3}.

It goes without saying that the calculation of 
higher order corrections in gauge theories 
is almost impossible without intensive use of 
computer algebra methods.  
In addition to the old problem of    
taking long traces of Dirac $\gamma$ 
matrixes, the algorithm of integration by parts,
when  applied even to a single 
3-loop p-integral, generically 
produces  dozens or even hundreds of  
terms.  At the moment there exist  
essentially three different packages 
which implement the algorithm. 
For p-integrals they are written  
in {\rm SCHOONSCHIP}
\cite{Veltman.2loopTadpole} 
(see Refs.~\cite{mincer1a,mincer1b} and in FORM 
\cite{Ver91} (see Refs.~ \cite{mincer2,mincer2S}).
\section{ \large\bfseries Infrared Rearrangement and $R^*$-operation}
At the moment there is no any general algorithm allowing to
analytically compute arbitrary 4-loop p-integrals. The problem of
creating such an algorithm seems to be hopelessly difficult. See
Ref.~\cite{aip2} where the point is dealt with  in some detail and also
a discussion in the conclusion of the present talk.

Nevertheless, a large group of important 4-loop problems ---
calculation of RG-functions or, equivalently, UV renormalization
constants --- proves to be reducible to the 3-loop case and, thus, be
doable in a completely analytic way.  This is because in MS scheme any
UV counterterm is polynomial in momenta and masses \cite{Collins75}.
This observation was effectively employed in Ref.~\cite{Vladimirov80}
to simplify considerably the calculation of UV counterterms.  The
method was further developed and named Infrared Rearrangement (IRR) in
Ref.~\cite{me79}.  It essentially amounts to an appropriate
transformation of the IR structure of FI's by setting zero some
external momenta and masses (in some cases after some differentiation
is performed with respect to the latter).  As a result the calculation
of UV counterterms is much simplified by reducing the problem to
evaluating massless p-integrals.  The method of IRR was ultimately
refined and freed from unessential qualifications in
Ref.~\cite{me84}. The following Theorem has been proven there by the
explicit construction of the corresponding algorithm:
\vglue 0.2cm                       

\noindent
{\bf 
Any UV counterterm for any 
(h+1)-loop Feynman 
integral can be expressed in terms of pole and finite parts 
of some appropriately  constructed (h)-loop p-integrals.
}

\noindent

In many important cases one is interested only in the absorptive part
of massless two-point correlators.  In this case available theoretical
tools are enough to guarantee at least {\em in principle} the
analytical calculability of absorptive part of an arbitrary 4-loop
p-integral.  This is eventually due to the fact that the absorptive
part of a four-loop p-integral is fully expressible in terms the
corresponding four-loop UV counterterm along with some three-loop
p-integrals.  As a typical example we will  consider in
next section the ratio   
$   R(s) =\sigma_{\mbox{tot}}(e^{+} e^- \to \mbox{hadrons}) /
  \sigma(e^+ e^- \to \mu^+ \mu^-)
$ 
which is  essentially 
given by the absorptive part of the  vacuum polarization.

The  Theorem  coupled with the  the integration by parts method
solves  {\em at least in principle} the task
of analytical evaluation of RG functions and absorptive parts of 4-loop
p-integrals.
It should be noted that the $R^*$-operation is  essential
to prove  the  Theorem, though in most (but not in all) practical
cases one could proceed without it.  However, such a practice, in
fact,  forces  diagram-wise  renormalization mode, what, in turns,
brings down a  heavy penalty of manual  treatment of  hundreds of 
diagrams.

Indeed, in  genuine four-loop calculations the reduction to three-loop 
$p$-integrals  is far from being 
trivial and  includes a lot of  manipulations.
Typical steps  here are
%\begin{itemize}
\begin{description}
\item[a]
to reduce the initial  Feynman integral to logarithmically divergent ones via
a proper differentiation with respect  masses and
external momenta;
\item[b]
 to identify  UV
and IR divergent subgraphs of the resulting integral;
\item[c]
to remove  in a recursive way 
the corresponding UV and IR divergences;
\item[d] to  compute   resulting p-integrals.
\end{description}
%\end{itemize}
Among these steps only the calculation of p-integrals can be at
present completely performed by a computer. All others, especially
{\bf b} and {\bf c} are difficult to computerize.  As a result, in
spite of the fact that the necessary theoretical tools have been
around for more than a decade, yet until very recently there existed
just one QCD four-loop result: the ${\cal O}(a_s^3)$ contribution to 
$R(s)$.  It was done in Refs.~\cite{GorKatLar91,SurSam91} within the
diagram-wise renormalization approach.  The core of the problem is the
calculation of the four-loop contribution to the photon anomalous
dimension entering into the RG equation for the photon polarization
operator. The initial 98 four-loop diagrams contributing to the photon
polarization operator proliferate to about 250 after the IRR procedure
is applied. In addition these diagrams contain about 600 various
subdiagrams which should be computed separately in order to subtract
UV subdivergences. No wonder that both calculations were  done
in  a particular gauge --- the Feynman one --- and, thus, a valuable 
possibility of  using the gauge invariance  as a  strong test of 
the correctness of the  results had not  been used at all.

The calculation of Ref.~\cite{GorKatLar91} did not use the
$R^*$-operation at all while that of Ref.~\cite{SurSam91} employed it
only for a few diagrams.  We will see in the next section that a
proper use of power of the $R^*$-operation allows one not only to
repeat with ease these calculations in the general covariant gauge but
also to extend them considerably.

%\newpage

\section{ \large\bfseries 4-loop case: photon vacuum polarization}
The photon vacuum polarization  function 
$\Pi(-q^2)$is defined through the correlator
of the hadronic EM current
$
j^{\rm em}_{\mu}=\sum_{{f}} Q_{{f}}
\overline{\psi}_{{f}} \gamma_{\mu} \psi_f
$
as follows 
\begin{equation}
\label{Pi}
\Pi_{\mu\mu}(q)  =
(4\pi)^2 i \int {\rm d} x e^{iqx}
\langle 0|T[ \;
\;j_{\mu}^{\rm em}(x)j_{\nu}^{\rm em}(0)\;]|0 \rangle
= 
\displaystyle
-3q^2\Pi(-q^2)
{}\, .
\end{equation}
The optical theorem
relates  the inclusive cross-section
and thus the function $R(s)$
to the discontinuity of $\Pi$
in the complex plane
\begin{equation} \label{d3}
R(s) =  \displaystyle
 \frac{3}{4\pi} \,{\rm Im}\, \Pi( - s -i\delta)
\label{discontinuity}
{}\, .
\end{equation}

The renormalization mode of the polarization operator 
$\Pi(Q^2)$  reads  
\beq
\Pi(Q^2/\mu^2,\alpha_s) = Z^{\rm em}  + \Pi_0(Q^2,\alpha_s^0),
{},
\label{renorm:mod}
\eeq
where $\alpha_s = g^2/(4\pi)$ is 
the  strong coupling constant.
Within the  
$\msbar$ scheme 
(here and  below  we are using a convenient combination $a_s = \alpha_s/\pi$)
\beq
Z^{\rm em} = \sum_{1\le j\le i} \left(Z^{\rm em}\right)_{ij} \frac{a_s^{i-1}}{\ep^j} 
{}\, ,
\label{Zem}
\eeq
with the coefficients $\left(Z_{\rm em}\right)_{ij}$ being pure numbers 
and $D=4-2\ep$ standing for the space-time
dimension. 

As a result we arrive at  the following renormalization group (RG) equation
for  the polarization operator  
($L_Q = \ln\frac{\mu^2}{Q^2}$) 
\beq
\frac{\prd }{\prd L_Q} \Pi =
 Q^2 \g_{\rm em}(a_s)
-\left(
 \beta( a_s) a_s\frac{\prd }{\prd a_s}
\right) \Pi
\label{rgPi2}
{}.
\eeq
Here
the photon anomalous dimension 
and the  $\beta(a_s)$-function  are  defined in the usual way
\begin{eqnarray}
\g_{\rm em} = \mu^2  \frac{d}{d \mu^2}(Z_{\rm em})-\epsilon
Z_{\rm em} =  
-\sum_{i\geq 0}(i+1)
(Z_{\rm em})_{i1}
a_s ^{i}
{} , 
\label{phpton:anom}
\\
\dmu a_s  = \as \beta(a_s) \equiv
-\sum_{i\geq0}\beta_i a_s^{i+2} 
\label{beta:def}
{}.
\end{eqnarray}
The relation (\ref{rgPi2}) explicitly demonstrates the main computational
advantage of finding first the polarization function $\Pi(Q^2)$
against a direct calculation of $R(s)$ in the case of massless
QCD
Indeed, in order $a_s^n$ the derivative $\frac{\prd }{\prd L_Q} \Pi$
and, consequently, $R(s)$ depends on the very function $\Pi$ which is
multiplied by at least one factor of $a_s$. This means that one needs
to know $\Pi$ up to order $a_s^{n-1}$ only to unambiguously
reconstruct all $Q$-dependent terms in $\Pi$ to order $a_s^n$,
provided, of course, the beta function and anomalous dimension
$\gamma_{\rm em} $ is known to order $a_s^n$. On the other hand, as we
have discussed before the calculation of an anomalous dimension or a
beta-function is known to be much easier than computing a correlator
of the same order in the coupling constant.

Thus, in order to check the results of
Refs.~\cite{GorKatLar91,SurSam91} by an independent calculation one
should compute $\Pi$ in order $a_s^2$ (a dozen three-loop diagrams)
and $\gamma_{\rm em}$ in order $a_s^3$ (more than a hundred of
four-loop diagrams). The first task is almost trivial nowadays (see the
previous two sections) so let us concentrate on the second one.  As we
have already discussed, the problem is doable via IRR and the
integration by parts method but the amount of
%human involvment in 
various non-trivial manipulations
with separate diagrams to be  performed by hand is  really terrific.

In a recent work \cite{gvvq} it has been demonstrated how the
formalism of the $R^*$-operation can be applied to derive an explicit
formula for the renormalization constant
$Z^{\rm em}$ and, consequently for $\gamma_{\rm em}$. 
The formula reads:
\begin{eqnarray}
&Z^{\rm em}& = - K_\ep \left\{  
%\left(
%
\frac{1}{Z_2}
\frac{1}{4 D}Tr[\delta\G^0_{\tilde\alpha}(0,0,a^0_s)\g_\alpha ] Z^{\rm em}
%\right)
-
\frac{\delta Z_2 }{Z_2} Z^{\rm em}
%%%%%%%%%%%%%%%%
%
\right.
\label{final_eq}
\\
&-& 
\left.
\frac{1}{2 D(D-1)} 
\Box_q Tr[\g_{\tilde \alpha} G^0(p+q,a^0_s)\G^0_\alpha(p,q,a^0_s) 
G^0(p,a^0_s)]|_{{}_{\displaystyle q=0}}
\right\}
\nonumber
{}\, .
%\label{final_eq}
\end{eqnarray}
Here $G^0$ and $\G^0_\alpha$ are the full quark propagator and the EM
current vertex function respectively; the integration with respect to
the loop momentum $ p $ with the weight function $\frac{
(4\pi)^2}{(2\pi)^D}$ is not explicitly displayed.
$Z_2$ is the quark wave function renormalization constant.

Several extra comments are in order regarding this formula.
\noindent
First,  Eq.~(\ref{final_eq}) is,  rigorously speaking,
applicable as it stands  only to the so-called non-singlet diagrams, that
is to those where both EM currents belong to one and the same quark
loop.  The four singlet diagrams, violating this requirement, appear
first  in  order $a_s^3$ and should be treated separately.

Second, by $\delta\G^0_{\tilde\alpha}(p,q,a^0_s)$
we denote the vertex function of the electromagnetic current
with the tree contribution removed. The ``tilde'' atop the index 
$\alpha$  again  that  in every diagram the quark propagator 
entering to the vertex $j_\alpha$ is softened at small momenta by
means of the auxiliary mass $m_0$  according to the rule
\beq
\g_\alpha  \to \g_{\tilde \alpha} = \g_\alpha \   p^2/({p^2-m_0^2})
{}.
\label{gamma_tilde} 
\eeq
The bare coupling constant  $a_s^0$ is to be understood as 
$a_s = Z_a a_s$,  with  $Z_a$ being 
the coupling constant renormalization constant.  

An inspection of (\ref{final_eq}) immediately shows that,
in order to find the ${(n+1)}$-loop  correction to $Z^{\rm em}$, 
one  needs only to  know the renormalization constants $Z_2$ and
$Z^{\rm em}$  to order $a_s^n$ as well as    the bare Green functions 
\beq
 G^0(p,a^0_s), \ \ 
\frac{\partial}{\partial q_\beta }
[\G^0_\alpha(p,q,a^0_s)]|_{{}_{\displaystyle q=0}} 
, \ \ 
\Box_q [\G^0_\alpha(p,q,a^0_s)]|_{{}_{\displaystyle q=0}}, 
\ \ 
\delta\G^0_{\tilde\alpha}(0,0,a^0_s)
\label{functions}
{}
\eeq
up to (and including) $n$-loops,  that is to order  $(a_s^0)^n$.

We have computed with the program MINCER 
\cite{mincer2}
the unrenormalized three-loop
Green functions (\ref{functions}) as well as the quark wave function
renormalization constant $Z_2$ to order $a_s^3$.  The calculations
have been performed in the general covariant gauge with the gluon
propagator $(g_{\mu\nu} - \xi \frac{q_\mu q_\nu}{q^2})/q^2$.  We have
also taken into account the singlet diagrams as well as extra diagrams
with some of virtual quarks replaced by colour octet neutral fermions.
In the minimal supersymmetric standard model such a fermion known as
gluino appears as the superpartner of the gluon
\cite{Farrar78}.  

Our  result for 
 $R(s)$ with $\mu^2 = s$ reads  \cite{gvvq}
\begin{equation}
%\begin{array}{l}
\dsp
R(s) = 3\sum_f Q_f^2\left\{1
+ 
a_s 
+
a^2_s 
r_2 
+
a^3_s 
 r_3
\right\}
%\\ 
\dsp
+ a_s^3
\left(\sum_f Q_f \right)^2
r_3^S
{},
\label{RS0}
\end{equation}
where $r_3^S {=}
\left(
\frac{55}{72} - \frac{5}{3}\zeta(3)
\right)
      $ and 
\begin{eqnarray}
r_2 &=& 
\frac{365}{24} 
-11  \,\zeta(3)
-\frac{11}{12}  \,n_f 
+\frac{2}{3}  \,\zeta(3) \,n_f 
-\frac{11}{4}  \,n_{\tilde g}
+2  \,\zeta(3) \,n_{\tilde g}
\label{r2}
{},
\nonumber\\
r_3 &{=}&
\frac{87029}{288} 
-\frac{121}{48}  \pi^2
-\frac{1103}{4}  \,\zeta(3)
+\frac{275}{6}  \,\zeta(5)
-\frac{7847}{216}  \,n_f 
+\frac{11}{36}  \pi^2 \,n_f
 \nonumber 
\\ {} 
%+\frac{11}{36}  \pi^2 \,n_f 
&+&\frac{262}{9}  \,\zeta(3) \,n_f 
-\frac{25}{9}  \,\zeta(5) \,n_f 
+\frac{151}{162}  \, n_f^2
-\frac{1}{108}  \pi^2 \, n_f^2
-\frac{19}{27}  \,\zeta(3) \, n_f^2
%\label{r3}
\nonumber 
\\ {} 
%-\frac{19}{27}  \,\zeta(3) \, n_f^2
&-&\frac{32903}{288}  \,n_{\tilde g}
+\frac{11}{12}  \pi^2 \,n_{\tilde g}
+\frac{277}{3}  \,\zeta(3) \,n_{\tilde g}
-\frac{25}{3}  \,\zeta(5) \,n_{\tilde g}
+\frac{151}{27}  \,n_f  \,n_{\tilde g}
 \nonumber \\ {} 
%+\frac{151}{27}  \,n_f  \,n_{\tilde g}
&-&\frac{1}{18}  \pi^2 \,n_f  \,n_{\tilde g}
-\frac{38}{9}  \,\zeta(3) \,n_f  \,n_{\tilde g}
+\frac{151}{18}  n_{\tilde g}^2
-\frac{1}{12}  \pi^2 n_{\tilde g}^2
-\frac{19}{3}  \,\zeta(3) n_{\tilde g}^2
\nonumber
%zero == 0
\rule{.0mm}{0.5cm}
{}.
\label{r3S}
\end{eqnarray}
Here $n_{\tilde g}$ is the number of neutral gluino multiplets  which we
take either zero or one. 
We observe that neither $\g_{em}$ no $R(s)$ depend on the gauge fixing
parameter $\xi$ as it must be. If $n_{\tilde g}$ is set to zero
then $R(s)$ is in complete agreement with the results of 
Refs.~\cite{GorKatLar91,SurSam91}.  
In the numerical form, 
\bea
&R(s)& = 3\sum_f Q_f^2 \left\{
1 + a_s +
 a_s^2 \left( 1.98571 - 0.115295 n_f - 0.345886 n_{\tilde g}
\right)
\right.
\nonumber
\\
&{}&
+ a_s^3
\left(
-6.63694 - 1.20013 n_f - 0.00518 n_f^2  - 2.85053 n_{\tilde g} 
\right.
\nonumber
\\
&{}&
\left.
\left.
%\rule{0.8cm}{0cm}
-0.03107 n_f n_{\tilde g} - 0.04661 n_{\tilde g}^2 
\right)
\right\}
\nonumber
%\\
%&-&
-\as^3 \left(\sum_f Q_f\right)^2
1.2395
{}.
\label{Rnum1}
\eea
\section{\large\bfseries   ${\cal O}(\as^3)$ gluino contribution to 
\\
$\Gamma_{tot}(Z \to \mbox{h's})$
and $\Gamma (\tau^- \to \nu_\tau + \mbox{h's})$}

The result \re{RS0} can be straightforwardly applied to find the 
${\cal O}(\alpha_s^3)$ gluino contribution 
to the $Z$-boson decay rate into
hadrons ($\Gamma^h_Z$).  As is well-known this decay rate may be
viewed as an incoherent sum of vector ($\Gamma^V_Z$) and axial
($\Gamma^A_Z$) contributions.  For massless $u,d,s,c$ and $b$ quarks
the QCD corrections to $\Gamma^V_Z$ are in one-to-one correspondence
to those for $R(s)$. If one neglects as we do\footnote{ These terms
have been found to be extremely small in Ref.~\protect{\cite{timo2}}}
the power suppressed terms of order $M_Z^2/(4M_t^2)$ and higher the same
statement is valid also for $\Gamma^A_Z$ except for a specific subset
of so-called singlet diagrams.  As a result 
the total decay rate is naturally 
presented as a sum of three
terms
\beq
\Gamma^{A,S}_Z = \Gamma^V_Z +\Gamma^{A,NS}_Z  + \Gamma^{A,S}_Z
{}.
\label{decomp}
\eeq
Here $\Gamma^{A,S}_Z$ and $\Gamma^{A,NS}_Z$ stand for the
contributions to $\Gamma^{A}_Z$ due to singlet and the rest
(non-singlet) diagrams respectively.  The first two terms in
(\ref{decomp}) are directly expressed 
through the coefficients $r_1 - r_3$ and $r_3^S$ as follows:
\[
\frac{\dsp \Gamma^V_Z +\Gamma^{A,NS}_Z}{\dsp \Gamma_0}
=
\left\{
 3 \sum_f (v_f^2+a_f^2)
\left[
1 + a_s + a_s^2 r_2 + a_s^3 r_3
\right]
+ 
\left(
\sum_f  v_f
\right)^2 r_3^S
\right\}
{},
\nonumber
%\label{Zboson1}
\]
where  
$\Gamma_0=G_{{\rm F}} M_{{Z}}^3/24 \sqrt{2} \pi=82.94$ 
MeV,
$v_{{f}} = 2 I^{{f}}_3 - 4Q_{{f}}
\sin^2 \theta_{{\rm w}}, a_{f} = 2 I^f_{{3}}$,
$a_s = a_s(M_Z)$ and we have  set $\mu = M_Z$.

The singlet diagrams are sensitive to the huge mass splitting in the
top-bottom doublet and should be computed afresh. To order $\alpha_s^2$
this was done in Ref.~\cite{KniKue90ab} 
(even without neglecting the power
suppressed terms) while the ${\cal O}(\alpha_s^3)$ corrections were
computed in \cite{ChetTar93}.  In fact, no extra calculations are
necessary to get the gluino contribution.  A simple inspection of
relevant diagrams immediately reveals that gluinos appear exclusively
through the one-loop fermion correction to a gluon propagator
This means that the 
result in QCD with gluinos can be obtained from the
one in pure QCD by the replacement 
$  n_f \to  n_f + 3{n_{\tilde g}} $. 
It  reads 
\bea
 \Gamma^{A,S}_Z &=& \Gamma_0\left\{
a_s^2
\left[
-\frac{37}{4} + 3\lgm \frac{M_Z^2}{M_t^2}
\right]
\right.
\nonumber
\label{Zboson2}
\\
&+&
a_s^3
\left[
-\frac{5825}{72}  + \frac{11\pi^2}{4}
+ 3\zeta(3)
 + \frac{19}{2}\lgm \frac{M_Z^2}{M_t^2}
 + \frac{33}{4}\lgm^2 \frac{M_Z^2}{M_t^2}
\right.
\\
&+&
\left.
\left.
\left(
n_f + 3 n_{\tilde g}
\right)
\left(
\frac{25}{12}  - \frac{\pi^2}{6}
+\frac{1}{3}\lgm \frac{M_Z^2}{M_t^2}
-\frac{1}{2}\lgm^2 \frac{M_Z^2}{M_t^2}
\right)
\right]
\right\}
\nonumber
{},
\eea
where  $M_t$ is the pole mass of the top quark and $n_f=5$
is the number of (light) quarks flavours. 
Numerically, the gluino term is tiny:
the above  equation with $n_f=5$, 
and $ M_Z^2 /M_t^2 = (91.187/180)^2$
is evaluated to 
\beq
\Gamma^{A,S}_Z = \Gamma_0
\left\{
-13.33 a_s^2 
+ a_s^2 
\left[
-52.514 
-2.8197 n_{\tilde{g}}  
\right]
\right\}
{}.
\label{GammaZsinglet:numerics}
\eeq

Another obvious application of  Eq.~\re{RS0} is the gluino  
contribution to the semihadronic decay rate
of the $\tau$-lepton. As is well-known in the massless limit
the perturbative contributions  to  the ratio
$
\displaystyle 
R_\tau = 
\frac{\G (\tau \to \nu_\tau +\mbox{hadrons})}%
{\G (\tau \to \nu_\tau e^- \bar{\nu_e})}
$
can be written as follows 
(for a  review including  non-perturbative
and other effects see Ref.~\cite{tau3})
\beq
R_\tau =  2\int_0^{M_\tau^2} \frac{\mathrm{d} s}{M_\tau^2}
(1-s/M_\tau^2)^2 (1 + 2s/M_\tau^2) \tilde{R}(s),
\label{Rtau}
\eeq
where $M_\tau^2$ is the $\tau$-lepton mass.
Here $\tilde{R}(s)$ is given by the expression \re{RS0} for $ R(s)$ with 
$3\sum Q_f^2$
replaced by $ 3(|V_{ud}|^2 +|V_{us}|^2)$, $n_f =3$ 
and $(\sum Q_f)^2$ set to zero.  After a straightforward
integration with respect to $s$ in \re{Rtau} one arrives at 
(for gluino in the octet representation)
\bea
R_\tau = 3 (|V_{ud}|^2 &+& |V_{us}|^2)
\left\{1
\right.
+  a_s 
+  a_s^2
\left[
5.2023 - 1.13755 n_{\tilde{g}}   
\right]
\nonumber
\\
&+&  a_s^3 
\left.
\left[
 26.366 - 21.0358 n_{\tilde{g}} 
+  1.42119 n_{\tilde{g}}^2
\right]
\right\}
{},
\label{tau:num}
\eea
with $a_s = a_s(M_\tau)$. 
\ice{
In[8]:= 
<<tau.m 

1. + ast + 5.20232 ast  + 26.3659 ast  - 1.13755 ast  ng - 21.0358 ast  ng + 
 
               3   2
>   1.42119 ast  ng

In[9]:= 
       }

Note that  the result \re{tau:num} 
is applicable only if the gluino mass
is significantly lower than that of the $\tau$-lepton.  
Taken at its face value  it could  spoil  the current
agreement between  the values of $\Lambda_{QCD}$ extracted from 
the $R_\tau$  and $\Gamma( Z  \to \mbox{hadrons})$.
Such a conclusion  seems to be premature
as  for a meaningful phenomenological discussion of the
${\cal O}(\alpha_s^3)$ gluino contributions  one should
also take into account the running of the coupling constant in the
next-next-to-leading order.  This requires the knowledge of the gluino
contribution to the three-loop coefficient $\beta_2$ which
is not yet available in the literature.

\section{\large\bfseries  Conclusion}
The skillful use of the $R^*$-operation leads to significant
simplifications of analytical 4-loop calculations if we are interested
in the divergent part of 4-loop diagrams.  Thus,  4-loop
calculations in QCD are gradually becoming practically feasible. 

It is only natural to ask now about next level, that is about
possibility to analytically compute {\em finite} parts of 4-loop
p-integrals and consequently (according to the Theorem from Section 3)
divergent parts of 5-loop p-integrals.  Even a superficial glance at
the problem shows that it hardly could be done completely
``analytically''. Still, there is, to my opinion, a chance that in a
sense the problem could be solved. I mean that a better understanding
of all kinds of identical relations connecting various
p-integrals\footnote{ For recent developments in this direction see
\cite{Bai96}.} could eventually result to the reduction of an
arbitrary 5-loop p-integral to a combination of some limited number (a few
dozens?)  of master p-integrals.  Once it is done there should be not
very difficult to evaluate the latter analytically or  numerically
with sufficiently high accuracy.
A  fresh example of such approach has been recently demonstrated 
in a work by   Laporta and Remiddi who achieved a computer-algebraic
reduction and eventual analytical calculation of all 3-loop 
diagrams contributing to the electron's anomalous 
magnetic moment \cite{Remiddi96}.

The present work has been submitted to the HEP-PH e-print archive
today, that is 29.10.1996. A related article \cite{levan:gluino} was
posted on the same archive yesterday.  Our results  for     the    
${\cal O}(\alpha_s^3)$ gluino contributions to $R(s)$ and to related
quantities have been confirmed there by an independent calculation.
\vskip0.3cm

\noindent
{\large\bfseries Acknowledgments }
\vskip0.3cm

I would like to thank S. Jadach for the occasion to present the
talk  at the Cracow International Symposium on
Radiative Corrections, 1-5 August 1996.

\sloppy
              \raggedright

\def\app#1#2#3{{\it Act. Phys. Pol. }{\bf B #1} (#2) #3}
\def\apa#1#2#3{{\it Act. Phys. Austr.}{\bf #1} (#2) #3}
                        \def\lhc{Proc. LHC Workshop, CERN 90-10}
\def\npb#1#2#3{{\it Nucl. Phys. }{\bf B #1} (#2) #3}
\def\plb#1#2#3{{\it Phys. Lett. }{\bf B #1} (#2) #3}
\def\prd#1#2#3{{\it Phys. Rev. }{\bf D #1} (#2) #3}
     \def\pR#1#2#3{{\it Phys. Rev. }{\bf #1} (#2) #3}
\def\prl#1#2#3{{\it Phys. Rev. Lett. }{\bf #1} (#2) #3}
\def\prc#1#2#3{{\it Phys. Reports }{\bf #1} (#2) #3}
\def\cpc#1#2#3{{\it Comp. Phys. Commun. }{\bf #1} (#2) #3}
\def\nim#1#2#3{{\it Nucl. Inst. Meth. }{\bf #1} (#2) #3}
\def\pr#1#2#3{{\it Phys. Reports }{\bf #1} (#2) #3}
\def\sovnp#1#2#3{{\it Sov. J. Nucl. Phys. }{\bf #1} (#2) #3}
\def\jl#1#2#3{{\it JETP Lett. }{\bf #1} (#2) #3}
\def\jet#1#2#3{{\it JETP Lett. }{\bf #1} (#2) #3}
\def\zpc#1#2#3{{\it Z. Phys. }{\bf C #1} (#2) #3}
\def\ptp#1#2#3{{\it Prog.~Theor.~Phys.~}{\bf #1} (#2) #3}
\def\nca#1#2#3{{\it Nouvo~Cim.~}{\bf #1A} (#2) #3}
\def\mpl#1#2#3{{\it Mod. Phys. Lett.~}{\bf A #1} (#2) #3}

%%%%%%%%%%%%%%%%%%%%%%%%%%%%%%%%%%%%%%%%%%%%%%%%%%%%%%%%%%%%%%%%%%%%%%%%

\end{document}